\documentclass[preprint]{aastex61}

\usepackage{CJK}
\usepackage{subfigure}
\usepackage{float}
\usepackage{color}
\usepackage{hyperref}
\usepackage{amsmath}

\definecolor{myred}{RGB}{255,66,56}                 
\definecolor{myblue}{RGB}{0,0,255}

\def\HIP   {$Hipparcos$}

\def\vxsgr   {VX Sgr}
\def\Jzeroei  {J1808$-$2124}
\def\Jtwoze   {J1820$-$2528}

\def\hho     {H$_2$O}

\def\VLSR    {$V_{\rm LSR}$}

\def\kms     {km~s$^{-1}$}
\def\masy    {mas~yr$^{-1}$}
\def\mjybeam {mJy~beam$^{-1}$}
\def\jybeam  {Jy~beam$^{-1}$}

\def\uas     {$\mu$as}
\def\Lsun    {$L_{\odot}$}
\def\Lbol    {$L_{\rm bol}$}
\def\dLsun   {$(d/{\rm kpc})^2~L_{\odot}$}

\def\Rsun    {$R_{\odot}$}

\def\h       {\ifmmode{^{\rm h}}\else{$^{\rm h}$}\fi}
\def\m       {\ifmmode{^{\rm m}}\else{$^{\rm m}$}\fi}
\def\s       {\ifmmode{^{\rm s}}\else{$^{\rm s}$}\fi}
\def\deg     {\ifmmode{^{\circ}}\else{$^{\circ}$}\fi}
\def\decdeg  {\ifmmode{{\rlap.}^{\circ}} \else ${\rlap.}^{\circ}$\fi}
\def\decs    {\ifmmode{{\rlap.}^{\rm s}} \else ${\rlap.}^{\rm s}$\fi}
\def\decas   {\ifmmode{{\rlap.}{''}}\else{${\rlap.}{''}$}\fi}

\def\Vsbar {\ifmmode {\overline{V_s}}\else {$\overline{V_s}$}\fi}
\def\Usbar {\ifmmode {\overline{U_s}}\else {$\overline{U_s}$}\fi}
\def\Wsbar {\ifmmode {\overline{W_s}}\else {$\overline{W_s}$}\fi}

\def\mux    {\ifmmode {\mu_x}\else {$\mu_x$}\fi}
\def\muy    {\ifmmode {\mu_y}\else {$\mu_y$}\fi}
\def\mura   {\ifmmode {\mu_{\alpha}}\else {$\mu_{\alpha}$}\fi}
\def\mude   {\ifmmode {\mu_{\delta}}\else {$\mu_{\delta}$}\fi}

\def\gax{\mathrel{\rlap{\lower4pt\hbox{\hskip1pt$\sim$}}
    \raise1pt\hbox{$>$}}}

\def\d    {\ifmmode {{\rlap{.}}^\circ}\else {${\rlap{.}}^\circ$}\fi}
\def\s    {\ifmmode {{\rlap{.}}^s}\else {${\rlap{.}}^s$}\fi}
\def\as   {\ifmmode {{\rlap{.}}^{''}}\else {${\rlap{.}}^{''}$}\fi}

\begin{document}

\begin{CJK*}{UTF8}{gbsn}

\title{The Parallax of the Red Hypergiant \vxsgr\ with Accurate Tropospheric Delay Calibration}

\author[0000-0003-2953-6442]{Shuangjing Xu (徐双敬)}
\affil{Shanghai Astronomical Observatory, Chinese Academy of Sciences\\
80 Nandan Road, Shanghai 200030, China}
\affil{University of Chinese Academy of Sciences, 19A Yuquanlu, Beijing
100049, China}

\author[0000-0003-1353-9040]{Bo Zhang (张波)}
\affil{Shanghai Astronomical Observatory, Chinese Academy of Sciences\\
80 Nandan Road, Shanghai 200030, China}

\author{Mark J.  Reid}
\affiliation{Harvard-Smithsonian Center for Astrophysics\\
60 Garden Street, Cambridge, MA 02138, USA}

\author{Karl M. Menten}
\affiliation{Max-Plank-Institut f\"ur Radioastronomie\\
Auf dem H\"ugel 69, 53121 Bonn, Germany}

\author{Xingwu Zheng (郑兴武) }
\affiliation{
School of Astronomy and Space Science, Nanjing University\\
22 Hankou Road, Nanjing  210093, China}

\author{Guangli Wang (王广利)}
\affil{Shanghai Astronomical Observatory, Chinese Academy of Sciences\\
80 Nandan Road, Shanghai 200030, China}
\affil{University of Chinese Academy of Sciences, 19A Yuquanlu, Beijing
100049, China}

\correspondingauthor{Bo Zhang}
\email{zb@shao.ac.cn}

\begin{abstract}

We report astrometric results of VLBI phase-referencing observations of
22 GHz \hho\ masers emission toward the red hypergiant \vxsgr, 
one of most massive and luminous red hypergiant stars in our Galaxy, 
using the Very Long Baseline Array. A background
source, \Jtwoze, projected 4\d4  from the target \vxsgr, was used
as the phase reference.  For the low declinations of these sources, 
such a large separation normally would seriously degrade the relative
astrometry. We use a two-step method of tropospheric delay calibration, 
which combines the VLBI geodetic-block (or GPS) calibration with an
image-optimization calibration, to obtain a trigonometric parallax of 
$0.64\pm0.04$ mas, corresponding to a distance of 1.56$^{+0.11}_{-0.10}$ kpc. 
The measured proper motion of \vxsgr\ is  $0.36\pm0.76$  and 
$-2.92\pm0.78$ \masy\ in the eastward and northward directions.
The parallax and proper motion confirms that \vxsgr\ belong to the Sgr OB1 association.
Rescaling bolometric luminosities in the literature to our parallax distance, 
we find the luminosity of \vxsgr~is $(1.95 \pm 0.62) \times 10^5$ \Lsun, where
the uncertainty is dominated by differing photometry measurements.
\end{abstract}

\keywords{ astrometry --- troposphere calibration --- masers ---
parallaxes --- proper motions --- stars:individual (VX Sgr) ---
stars:supergiants}

\section{INTRODUCTION}\label{sec:intro} 

VX Sagittarii (\vxsgr) is one of most massive and luminous red
hypergiant stars in our Galaxy. It is a semi-regular variable with a
period about 732 days~\citep{2017ARep...61...80S}. Distances reported
in the literature for VX Sgr range from 0.3 to 2.0 kpc
~\citep{2007ChJAA...7..531C}. The parallax measured by \HIP\ is 3.82
$\pm$ 2.73 mas~\citep{2007A&A...474..653V}, which really provides only weak
limits for its distance.  The first data release of {\it Gaia}
does not include a parallax for \vxsgr, and since red hypergiants are large, 
variable, and often surrounded by copious dust, these issues may limit the 
accuracy of future optical astrometry.  Of the ``fabulous-four'' 
red hypergiants in the Milky Way, that all show strong maser emission
 VLBI parallaxes have been determined for VY CMa~\citep{2012ApJ...744...23Z,2008PASJ...60.1007C}, NML Cyg~\citep{
2012A&A...544A..42Z} and S Per ~\citep{2010ApJ...721..267A}, leaving only
\vxsgr\ without a reliable distance.

Currently, parallaxes for sources across
the Milky Way are being measured with $\sim10$ \uas\ accuracy using
VLBI phase-referencing techniques.
Among the various error sources in VLBI phase-referencing, 
uncompensated tropospheric delays typically dominate systemic
errors at observing frequencies $\gax$10 GHz
~\citep{2014ARA&A..52..339R}.  Generally, tropospheric delay, $\tau$, 
can be modeled by a
zenith delay, $\tau_0$, multiplied by a factor (the mapping
function) of $\approx\sec{Z}$, where ${Z}$ is the local source zenith angle.
As described in \citet{1999ApJ...524..816R}, the difference in delay error 
for a single antenna when observing two sources separated in zenith
angle by $\Delta Z$ is $\approx\tau_0\sec{Z}\tan{Z}\Delta Z$.
Thus, large separations between the target and a calibrator
and low elevation observations can lead to large astrometric errors,
and a method to
calibrate the tropospheric delay is crucial to obtain accurate
parallax measurements.

In this paper, we present astrometric results from multi-epoch 
Very Long Baseline Array (VLBA)
observations of the 22 GHz \hho\ maser emission toward \vxsgr\ and 
continuum emission of extragalactic quasars.  
In \S\ref{sec:obs}, we describe the 
phase-referencing observations.  We
discuss methods for accurate troposphere calibration in \S\ref{sec:trops},
and a procedure to obtain residual tropospheric delay biases in \S\ref{sec:app}. 
In \S\ref{sec:para}, we use the time variation of the positions of maser
spots relative to a background source to determine a trigonometric
parallax and absolute proper motion with different troposphere
calibration methods.   Finally the future outlook of the
calibration method is discussed in \S\ref{sec:disc}.

\section{OBSERVATIONS} \label{sec:obs} 

We conducted multi-epoch VLBI phase-referencing observations of VX Sgr
at 22 and 43 GHz with the VLBA under program BZ039 with 7-hour tracks on
2009 September 17 (BZ039A), 2010 May 3 (BZ039C) and May 6 (BZ039B), and 
September 15 (BZ039D).  The Fort Davis antenna did not participate in 
program BZ039A and the Owens Valley antenna did not participate in program BZ039B. 
The observations sample near the peaks of the sinusoidal trigonometric 
parallax curve in the East-West direction, resulting in low correlation 
coefficients between the parallax and proper motion parameters.  
We observed two extragalactic radio sources as
potential background references for parallax solutions.  We alternated
between $\approx16$ min blocks at 22 and 43 GHz.  
The 43 GHz  SiO maser emission toward \vxsgr\ was weak, and phase referencing 
with the maser succeeded only for limited times for a few antennas. 
Therefore, we only used the 22 GHz data to obtain the parallax and do not
discuss the 43 GHz data in the rest of this paper.  Within a block
the observing sequence repeated VX Sgr, \Jzeroei, VX Sgr, \Jtwoze,
switching between the maser target and a background source every 40 s,
typically achieving 30 s of on-source data. 
Table~\ref{tab:src_pos} lists observational parameters for the sources.

\begin{deluxetable}{lrrrrrcc}
  \tablecolumns{8}
  \tablewidth{0pc}
  \tablecaption{Source Positions and Brightnesses at 22 GHz}
  \tablehead{
  \colhead{Source} & \colhead{R.A. (J2000)} & \colhead{Dec.  (J2000)}               & \colhead{$S_p$}     & \colhead{$\theta_{sep}$} & \colhead{P.A.}     & \colhead{\VLSR}  & \colhead{Beam} \\
    \colhead{      } & \colhead{(h~~~m~~~s)}  & \colhead{(\degr~~~\arcmin~~~\arcsec)} & \colhead{(Jy/beam)} & \colhead{(\degr)}        & \colhead{(\degr)}  & \colhead{(\kms)} & \colhead{(mas~~mas~~\degr)} \\
  }
  \startdata
  \vxsgr    & 18 08 04.0510  & $-$22 13 26.566  &   28$-$60   &  ...  & ... &  1.2    & 0.7 $\times$ 0.3 @ $-$5\\
  \Jzeroei  & 18 08 06.8471  & $-$21 24 45.128  &  $<0.05$   &  0.8  &  $-179$  & ...  & ... \\
  \Jtwoze   & 18 20 57.8487  & $-$25 28 12.584  &    0.39   &  4.4  &  $-43$  & ...  & 0.9 $\times$ 0.3 @ $-$7 \\
  \enddata
  \tablecomments{
  $S_p$ is the peak source brightnesses and
  \VLSR\ is the Local Standard of Rest velocity of the maser reference feature.  
  $\theta_{sep}$ and P.A. indicate source separations and position 
  angles (East of North) from \vxsgr.  
  The last column gives the FWHM
  size and position angle (East of North) of the Gaussian restoring beam.  Calibrator \Jtwoze~is
  from the ICRF~\citep{1998AJ....116..516M} and \Jzeroei~is from
  \url{http://astrogeo.org}. 
  }
  \label{tab:src_pos}
\end{deluxetable}

We placed observations of two strong sources 3C345 (J1642$+$3948) and
3C454.3 (J2253$+$1608) near the beginning, middle, and end of the
observations in order to monitor delay and electronic phase differences
among the intermediate-frequency bands.  The rapid-switching
observations employed four adjacent bands of 8~MHz bandwidth and
recorded both right and left circularly polarized signals.  The \hho\
 masers were contained in the second band centered at a \VLSR\ of
5 \kms.  In order to perform atmospheric delay calibration, we placed
``geodetic'' blocks before and after our phase-referencing
observations~\citep{2009ApJ...693..397R}.  These data were taken in left
circular polarization with eight 8~MHz bands that spanned 480~MHz 
between 22.0 and 22.5~GHz; the bands were spaced in a
``minimum redundancy configuration'' to uniformly sample frequency differences.

The data were correlated in two passes with the VLBA correlator in
Socorro, NM.  One pass generated 16 spectral channels for all the data
and a second pass generated 256 spectral channels, but only for the
single (dual-polarized) frequency band containing the maser signals,
giving a velocity resolution of 0.42 \kms.  
The data calibration was performed
with the NRAO Astronomical Image Processing System (AIPS), following
procedures described in \citet{2009ApJ...693..397R}.  We used
an \hho\ maser spot at  a \VLSR\ of 1.2 \kms\ as the phase-reference, 
because it was considerably stronger than the background sources and could be detected
on individual baselines in the available on-source time.

Unfortunately,  \Jzeroei, which is projected within 1\deg\ of \vxsgr, 
was too weak to be detected in our observations, and
we could only use \Jtwoze\ as a background reference to
determine the parallax. \vxsgr\ is $\approx$ 4\decdeg4\ away from
\Jtwoze, and, given the low declinations of the sources, they were observed at
elevation angles $<40$\deg\ at most of the VLBA antennas.  
These characteristics accentuate the need for accurate tropospheric delay calibration
for an accurate parallax measurement.

\section{ Accurate tropospheric delay calibration}
\label{sec:trops}

The atmospheric delay is almost entirely non-dispersive (frequency
independent) at frequencies above $\sim10$ GHz.  It is convenient to separate the
tropospheric delay into two components based on the timescales of
fluctuation: a slowly ($\sim$hours) and rapidly ($\sim$minutes)
varying term. The rapid tropospheric delay variation is calibrated by
observing the target and a nearby reference source simultaneously or
fast-switching in phase-referencing observations. 
The slowly varying tropospheric delays can be calibrated by geodetic-like VLBI observing blocks 
(``geodetic blocks'') or by using Global Positioning System (GPS)
data~\citep{2014ARA&A..52..339R}. For our observations, 
six of the ten VLBA stations were co-located with GPS
antennas, which offers an opportunity to calibrate tropospheric delay at those
stations using either method. 
In Table~\ref{tab:vlba_gps}, the 2-letter VLBA and 4-letter IGS acronyms
are given as well as the distance between the VLBI and GPS antennas and
their height differences (VLBI - GPS). The VLBA correlator model
removes the effects of a constant tropospheric zenith delay
using Saastamoinen's formulas and seasonally averaged parameters.
It is customary to separate the zenith
total delay (ZTD) into ``dry'' and ``wet'' components.  The
dry component can be accurately modeled from surface weather parameters, and therefore
residual zenith delays are assumed to be dominated by the zenith wet delays (ZWDs).
The residual ZWDs can be obtained from GPS ZTD products by removing 
the tropospheric delays in the VLBA correlator model, and correcting the
differences in the ZWDs owing to the different locations, and 
especially altitudes, of the VLBA and GPS antennas~\citep{2008ChJAA...8..127Z}.
Note that the VLBA geodetic blocks provide only two measures of the ZWD separated by
about six hours, whereas the GPS data from
the IGS (International GNSS Service) provides estimates every 5 minutes.

\begin{table}[H]
    \footnotesize
    \caption[]{VLBA Stations Collocated with GPS Antennas\label{tab:vlba_gps}}
    \begin{center}
        \begin{tabular}{cllrr}
            \hline \hline
            Code   &  Location       & GPS code & Separation (m)  &  Height Diff. (m)   \\
                   &                 &          &               &  (VLBI-GPS) \\
            \hline
            BR     &  Brewster       & BREW     &    59.5  & 11.9\\
            FD     &  Fort Davis     & MDO1     &  8417.6  & –398.1\\
            MK     &  Mauna Kea      & MKEA     &   87.8   & 8.4\\
            NL     &  North Liberty  & NLIB     &   67.3   & 15.2\\
            PT     &  Pie Town       & PIE1     &   61.8   & 17.0\\
            SC     &  Saint Croix    & CRO1     &   82.6   & 16.9\\
            \hline
        \end{tabular}
    \end{center}
\end{table}

Many authors have compared the ZTD estimated from GPS data
to geodetic VLBI observations and found that the differences are
very small (a few millimeters) when averaged over periods of weeks
to decades \citep{2007JGeod..81..503S,2012JGeod..86..565N,2013JGeod..87..981T}.
However, for time scales of hours, which is important for VLBI
phase-referenced imaging, while standard deviations can be at sub-cm levels, 
systematic offsets (biases) are sometimes greater than 2 cm \citep{2008ChJAA...8..127Z}.  
A zenith path-delay error of 2 cm can leads to 
a relative positional error $\sim100$ \uas\ for low declination sources 
($\delta \approx -30\deg$) based on simulations of VLBI
phase-referencing observations for source separations of 1\deg\  
\citep{2008PASJ...60..951H,2006A&A...452.1099P}.

\begin{figure}[H] 
    \centering
    \includegraphics[angle=-0,scale=0.4]{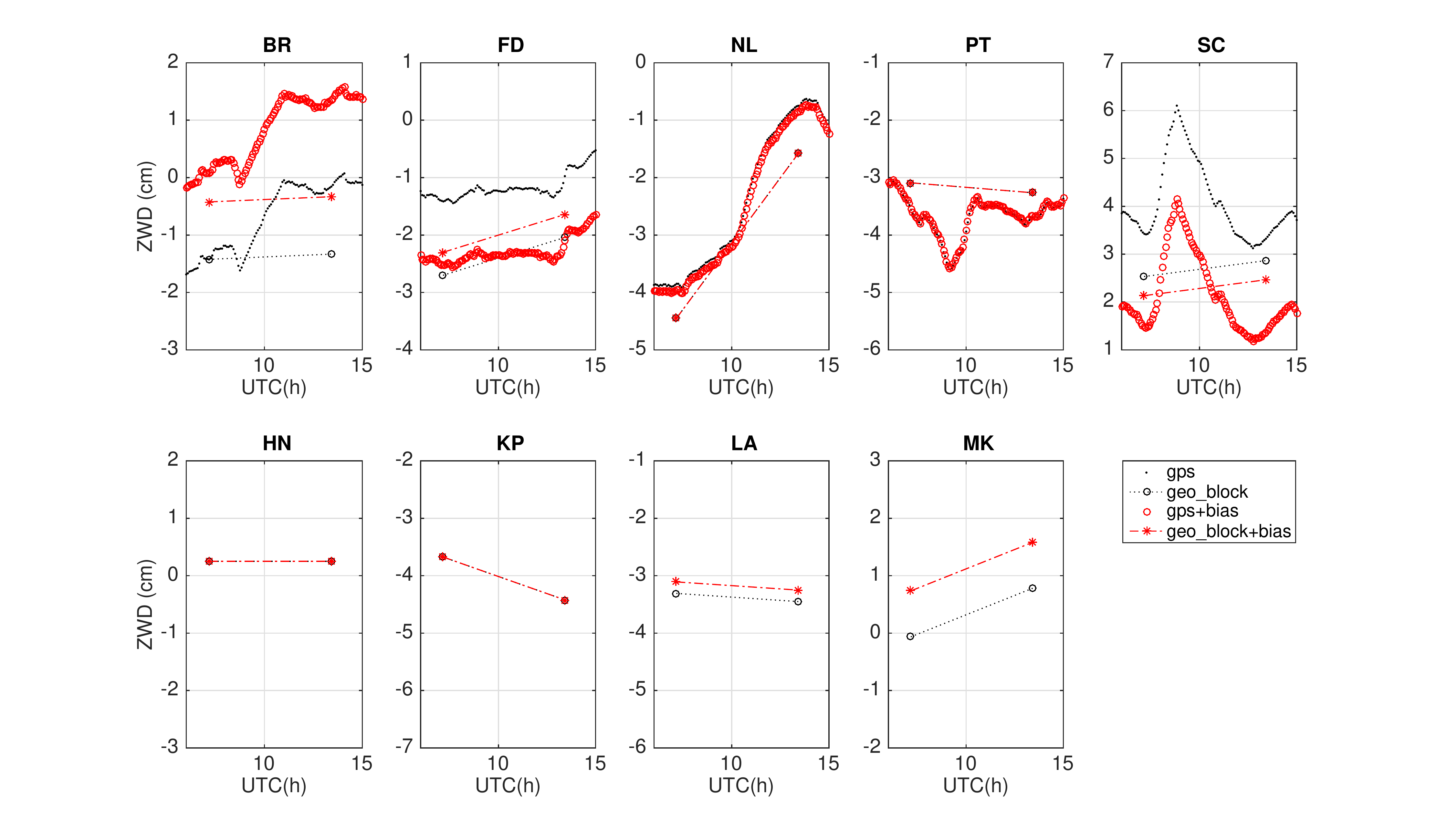}
    \caption{
    The residual zenith wet delay(ZWD) from GPS ({\it black dots}) and 
    geodetic-block ({\it black circles with dotted lines}) data at nine VLBA
    stations during observations for program BZ039B.
    Corresponding {\it red
    circles} and {\it red stars with dash-dotted lines} are ZWDs
    with estimated biases removed as described in
    \S\ref{sec:app}.
    }
    \label{fig:zwd}
\end{figure}

Figure~\ref{fig:zwd} shows a comparison between the residual ZWDs
from GPS and geodetic-block observations.
The Owens Valley (OV) antenna did not participate in this session and the GPS 
ZTD products were not available for the Mauna Kea (MK) station. The ZWDs between
the two geodetic-blocks are linear interpolations.
The systematic biases between GPS and geodetic-block ZWD data ({\it black markers}
in Figure~\ref{fig:zwd}) at stations FD and SC
are about 2 cm, which are consistent with those reported
by \citet{2008ChJAA...8..127Z} and \citet{2008PASJ...60..951H}.


\begin{figure}[H] 
    \centering

    \includegraphics[angle=0,scale=0.8]{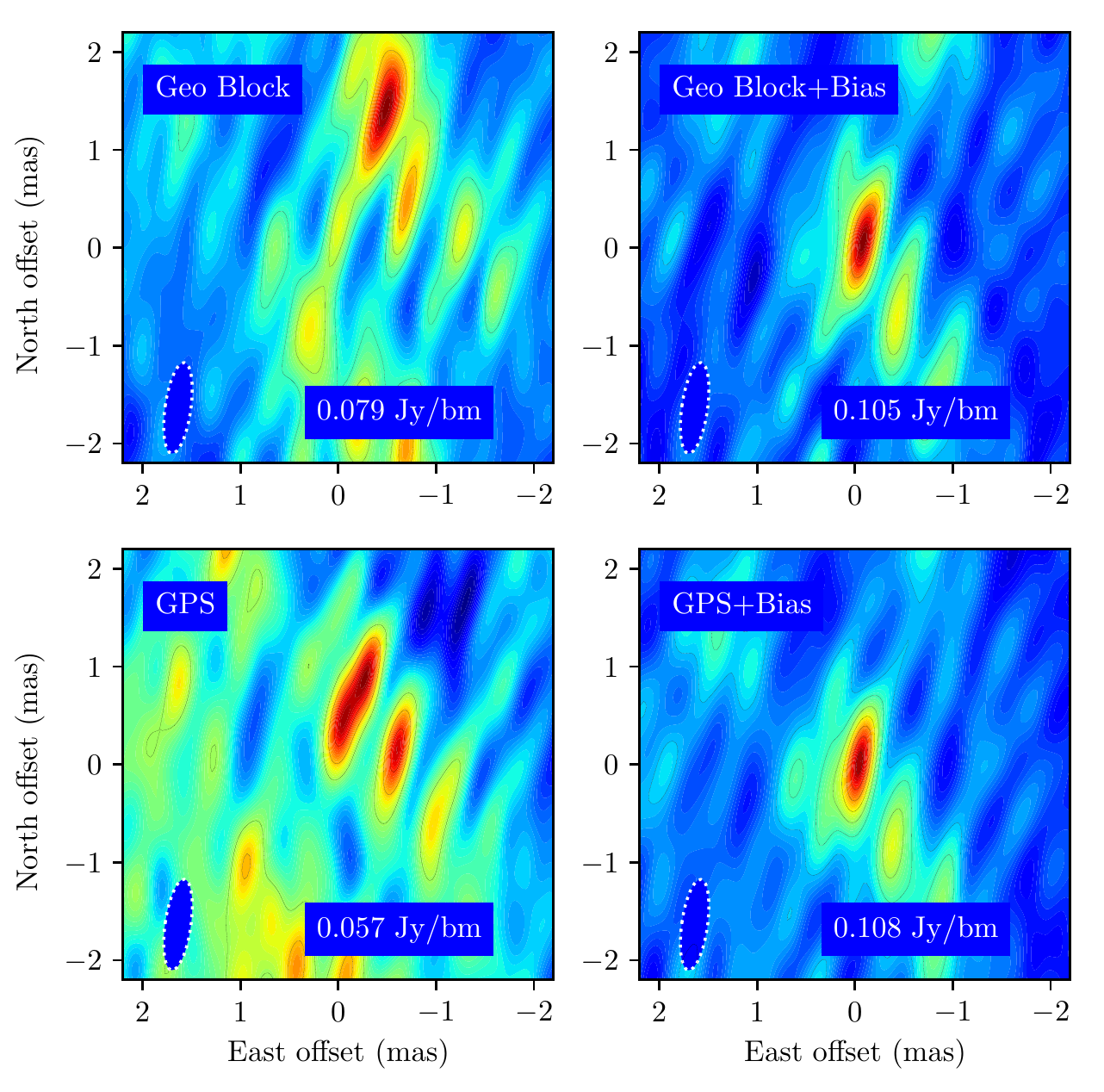}

    \caption{
    Phase-referenced images of \Jtwoze\ during the observation for program BZ039B, with geodetic-like observations
    ({\it upper
    left panel}), with geodetic-like observations plus bias correction
    ({\it upper right panel}),
    with GPS data ({\it bottom left panel}) and with GPS data plus bias
    correction ({\it bottom
    right panel}). Contour levels are spaced linearly at 15 \mjybeam.
    }
    \label{fig:img_qso}
\end{figure}

The {\it left panels} of
Figure~\ref{fig:img_qso} show the phase-referenced images of \Jtwoze~made from data
 that had tropospheric delays calibrated with geodetic-block observations (top),
with peak brightness of 79 \mjybeam, 
and GPS data (bottom), with peak brightness of 57 \mjybeam.
(Note, that when calibrating data with the GPS method, for the four VLBA stations 
without GPS systems we substituted the geodetic-block estimates of wet zenith delay.) 
The images suggest that the ZWDs from the geodetic block calibration
are better than for the GPS calibration. However, both
images appear confused with several probable spurious components, 
indicating that there may still remain systematic errors with both
methods. This motivated us to examine possible residual phase biases.

Two methods have been discussed in the literature to deal with
residual phase biases and both involve a second-step calibration,
either image optimization~\citep{2008PASJ...60..951H} or 
phase-fitting \citep{2005ASPC..340..455B}.  These calibrations 
work on the phase-referenced visibilities themselves, and
are considered {\it self-calibration} methods, as opposed to 
GPS or  geodetic block calibrations which use independent observations.
Both self-calibration methods assume constant zenith delay-errors (biases)
at each station over the observation period.  
The phase-fitting method is only appropriate for interferometer visibilities 
with high SNR, while the image optimization method can be applied to weak
sources~\citep{2008PASJ...60..951H}.  Thus, we explored the image
optimization approach here.   

\begin{figure}[H]
    \centering
    \includegraphics[angle=0,scale=0.41]{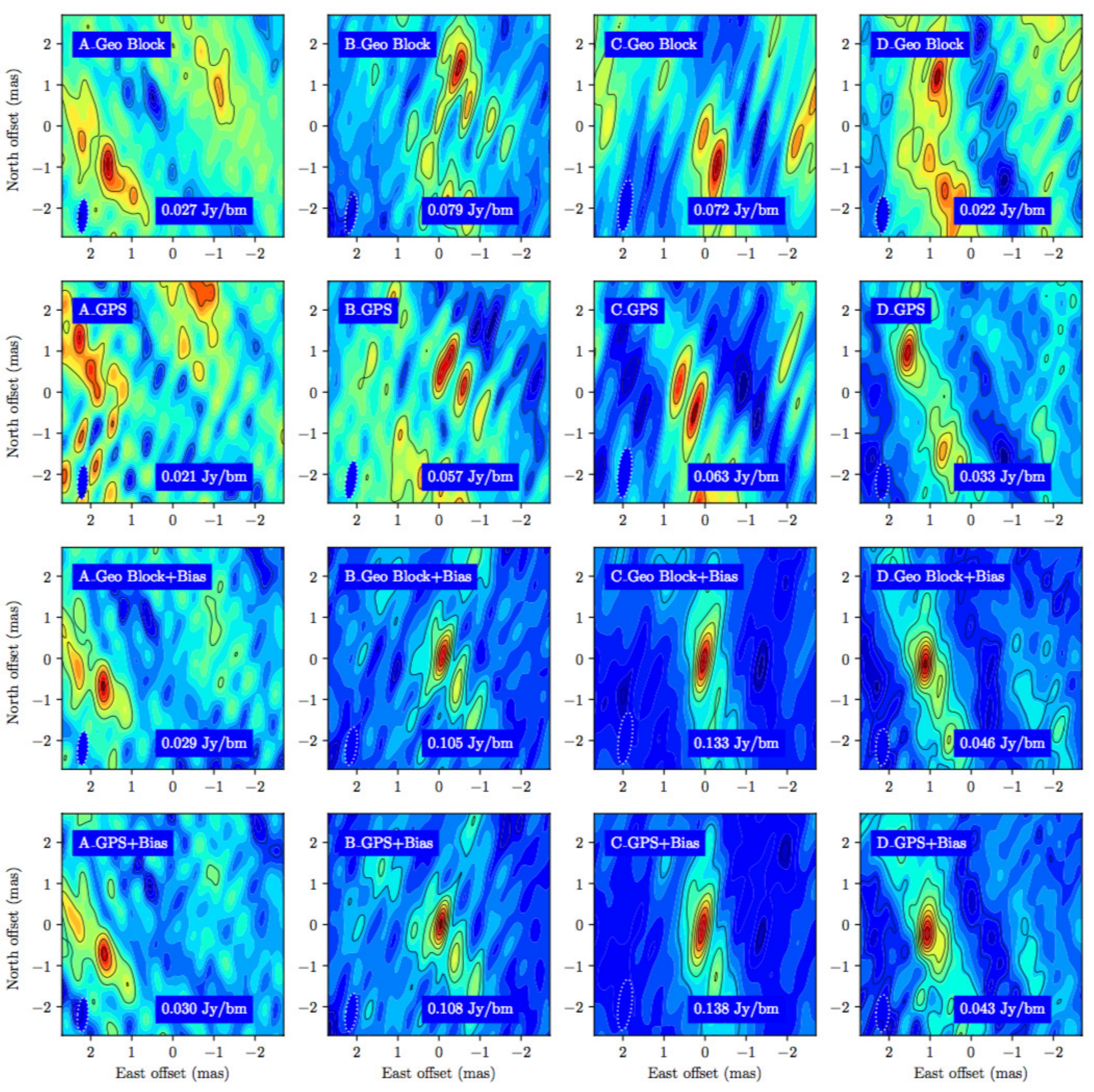}
    \caption{
    Phased-referenced image of \Jtwoze\  with different troposphere calibration methods. 
    Panels from top to bottom are for calibrations with
    geodetic-blocks, GPS, geodetic-blocks with bias correction, 
    and GPS with bias correction. 
	Panels from left to right are for the VLBA program BZ039A, BZ039B, BZ039C 
	 and BZ039D, labeled with A, B, C and D, respectively.
    Contour levels are spaced linearly at 7 \mjybeam\ (BZ039A), 15 \mjybeam\ (BZ039B), 20 \mjybeam\ (BZ039C) and 5 \mjybeam\ (BZ039D) , respectively.}
    \label{fig:qso_4}
\end{figure}

Following the process described in detail in \S~\ref{sec:app},
we obtained a residual (total zenith) phase bias for each VLBA station that had been
calibrated either by GPS data or geodetic blocks.
Residual tropospheric ZWDs with phase biases removed are shown in 
Figure~\ref{fig:zwd} with the {\it red markers}.  The phase biases 
correspond to about 1 cm of path delay for most stations.  The {\it right
panels} in Figure~\ref{fig:img_qso} show the images of
\Jtwoze\ with biases removed.  Compared to the images without biases removed,
the peak intensities increased from 79 to
105 \mjybeam\ for geodetic-block calibration, and from 57 to 108
\mjybeam\ for the GPS method.  Figure~\ref{fig:qso_4} provides images from
all four epochs for the different calibration methods. 
One can see that the bias-corrected images show simpler structure with
a brighter dominant component than uncorrected images. 
We will discuss the improvement of astrometric accuracy with the bias
correction in \S\ref{sec:para}.

\section{The method to obtain the bias}
\label{sec:app}

The image-optimization method for secondary calibration is described 
in~\citet{2008PASJ...60..951H}.   The method essentially is a 
brute-force technique in which a grid of trial zenith delays are
removed from the visibility data and subsequent maps are optimized 
for their peak brightnesses.  The basic idea is
that zenith delay errors degrade phase-referenced images, and the image
with the greatest peak brightness should indicate the best 
zenith delay calibration. 

As mentioned in \S\ref{sec:trops}, after primary calibration with either
GPS data or geodetic blocks, residual zenith delay errors of 
1 to 2 cm can exist.  For our study, trial zenith path-delays for each antenna
ranging from $\pm3$ cm, with steps of between 0.2 to 0.5 cm, were removed
from the data. The visibility data were then Fourier-transformed to make synthesized 
images, and the peak brightness is measured (without CLEAN deconvolution). 
In principle, it would be best to adjust residual biases simultaneously 
for all 10 VLBA antennas.  But, this would be very time consuming, since even a
10 trial by 10 station grid would require constructing $10^{10}$ 
phase-referenced maps. Instead, we adjusted the trial zenith delays
for one given antenna, holding the others constant, to achieve an optimal
value for that antenna. 

\begin{figure}[H]
    \centering
    \includegraphics[angle=0,scale=0.8]{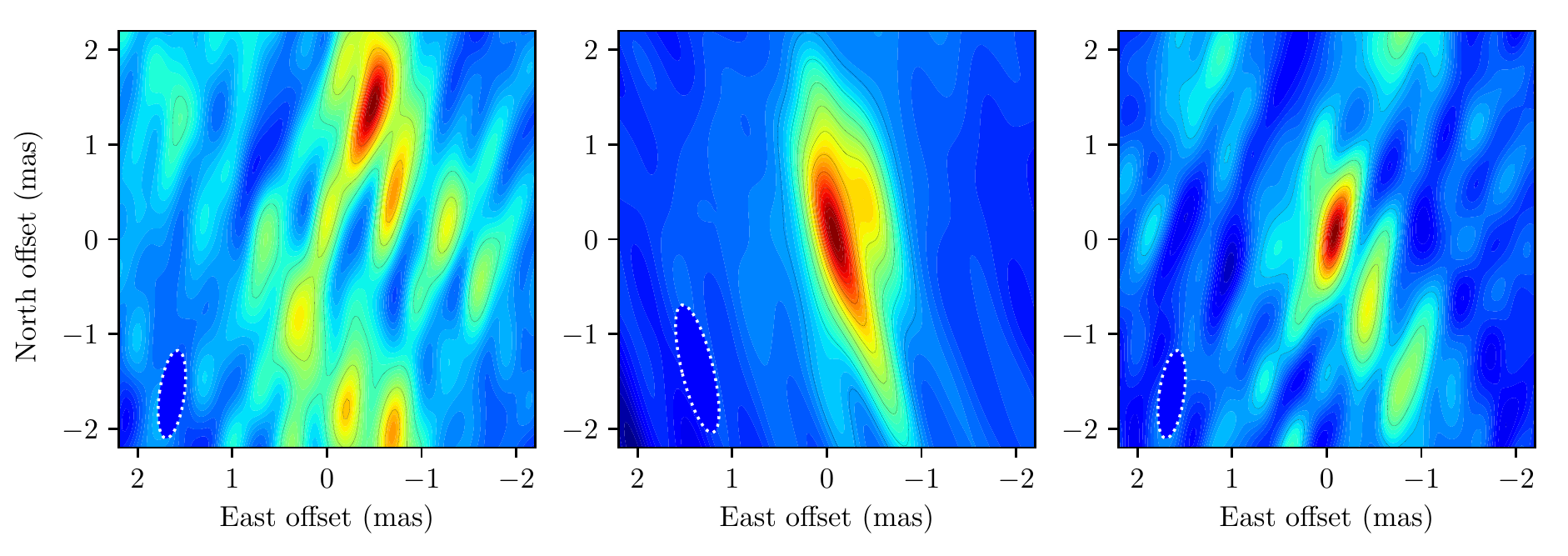}
    \caption{
    Phased-referenced images of \Jtwoze\  after the geodetic-block calibration for 
    the VLBA program BZ039B. {\it Left panel}: using all stations. {\it Middle
    panel}: removing the BR, MK and NL stations.
     {\it Right panel}: using all stations
    after the bias correction. Contour levels are spaced linearly at 15 \mjybeam\ .
    }
    \label{fig:opt}
\end{figure}

In our experience, the image-optimization method requires a reliable initial image. 
As shown in the left panel of Figure~\ref{fig:opt}, after the geodetic-block
calibration, the image made using data from all
the stations shows several components.
We cannot use the image-optimization method directly on such a poor image.
Removing different stations from the uv database, we
found that some stations degrade the image quality seriously (probably
owing to large phase biases). After removing the problematic stations, the image 
showed one clear dominant component (see the middle panel of Figure~\ref{fig:opt}).
Next we used the image-optimization method based on this single component
image to optimize the phase biases for the ``clean'' stations. 
After doing this, we introduced the ``problem'' stations one at a time and 
obtained preliminary phase biases for them.
Finally, we iterated the procedure with all stations.  The right
panel of Figure~\ref{fig:opt} shows the finial image with the bias
correction for all stations. 

\begin{figure}[H]
    \centering
    \includegraphics[angle=0,scale=0.65]{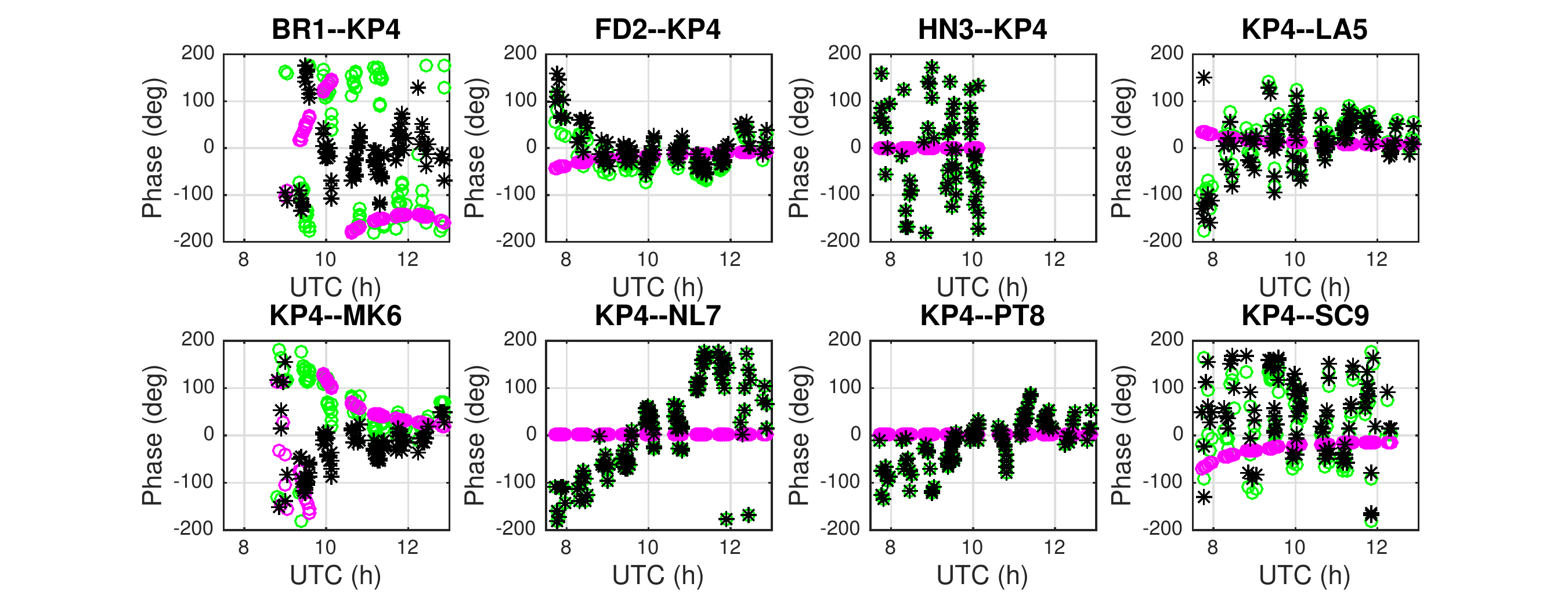}
    \caption{
    Interferometer phases for \Jtwoze\ for program BZ039B. 
    {\it Green dots}: data using the geodetic-block calibration; 
    {\it magenta dots}: model phases based on zenith biases;
    {\it black dots}: residual phases after bias correction. 
    }
    \label{fig:fit}
\end{figure}

In order to assess the image optimization process, we examined the residual phases on
each baseline.  We shifted the source position so that the brightest component
was near the map origin, which for perfect calibration should yield zero phases.
Figure~\ref{fig:fit} shows these phases (green dots) for all VLBA baselines to the
Kitt Peak (KP) station. The predicted phases based on the image optimization 
procedure (magenta dots for {\it zenith} phase shifts of BR = 1 cm, FD = 0.4 cm, 
LA = 0.2 cm, MK= 0.8 cm and SC = $-0.4$ cm) are also shown, as are residual 
differences (black dots).   
The BR and MK stations with the zenith phase biases of 1 cm show moderate phase
residuals.  The NL station residuals are quite large and systematic, possibly
indicating that the assumption of a constant zenith bias is not valid here.
Indeed, the ZWDs for NL changed by about 3 cm over the observations in BZ039B
(see Fig.~\ref{fig:zwd}), the most for any station.  Methods to address discrepant 
stations like NL remain to be developed.

\section{Parallax and Proper Motion}

\label{sec:para}

\begin{figure}[H]
    \centering
    \includegraphics[angle=0, scale=0.55]{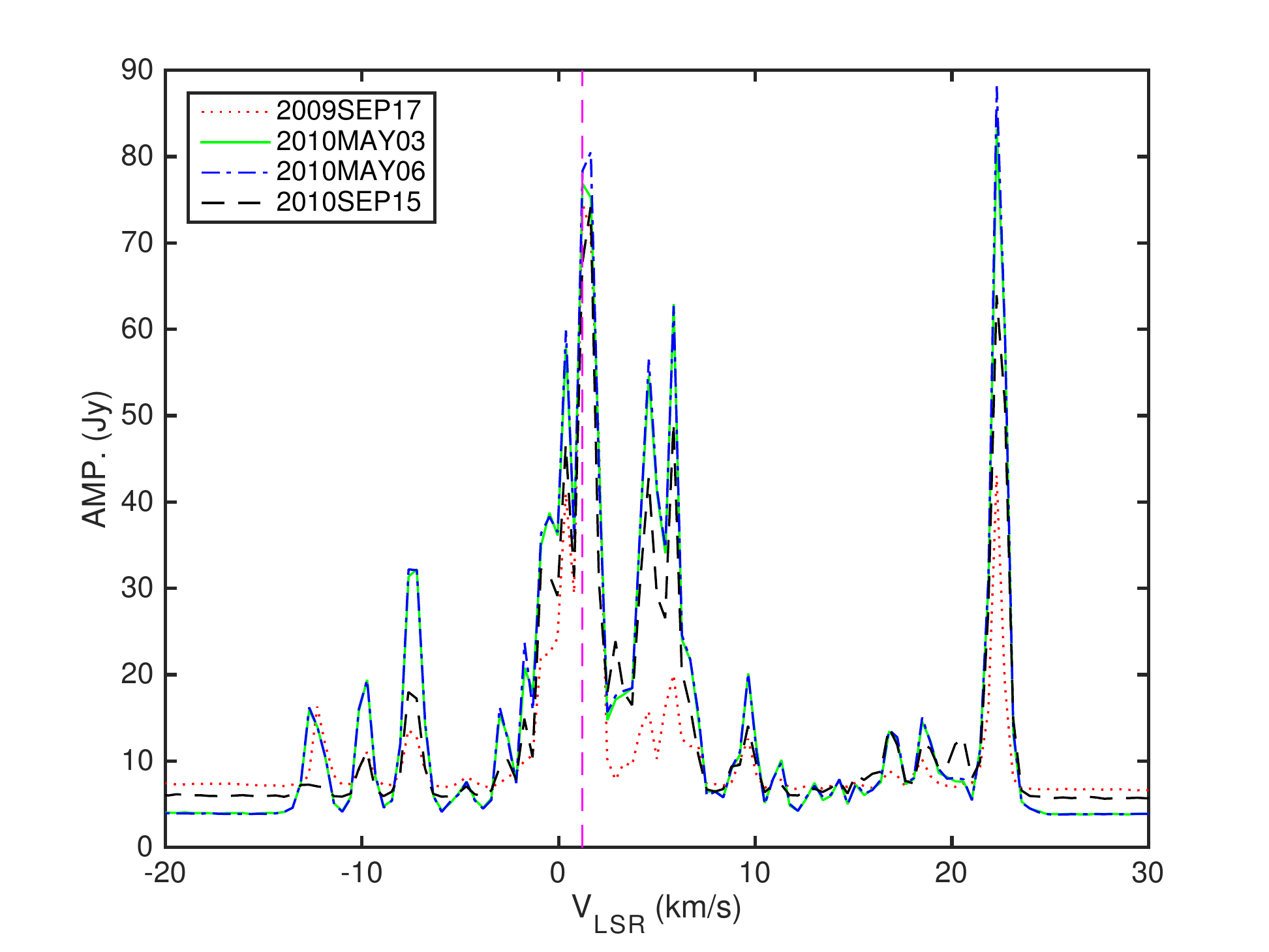}
    \caption{
    Interferometer spectra (scalar averaged cross-power amplitude over the full
     duration of the observation) of the \hho\ masers toward \vxsgr\ 
     from Kitt Peak to Los Alamos baseline at four 
     epochs. The red dashed line indicates the maser feature at VLSR of  
     \VLSR\ = 1.2 \kms\ which served as the interferometer phase reference. }
    \label{fig:line}
\end{figure}

\begin{figure}[H]
  \centering
\includegraphics[angle=0, scale=0.75]{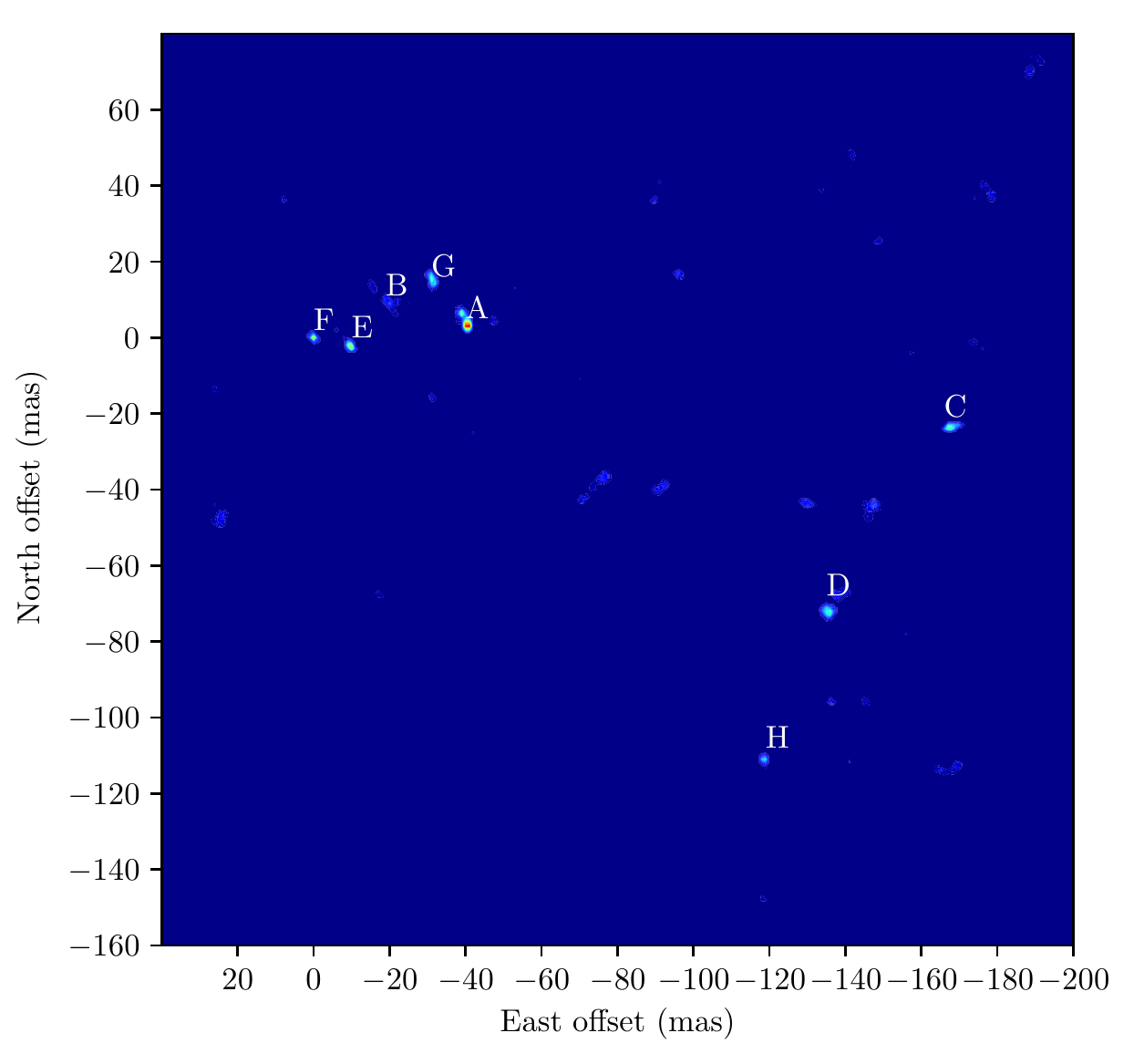}
  \caption{
Spatial distribution of the velocity integrated \hho\ maser features toward 
\vxsgr\ from VLBA observations on 2009 September 17.
}
\label{fig:maser}
\end{figure}

Figure~\ref{fig:line} shows the interferometer spectra of \hho\ maser 
emission observed with one VLBA baseline at all four epochs. 
The \hho\ maser emission spans a \VLSR\ range from about --15 to 25 \kms.  The
flux densities of maser features  between \VLSR\ of 0 and 10 \kms\ 
varied considerably from epoch to epoch, while the peaks of blue-shifted (\VLSR\
$\approx-10$ \kms) and red-shifted (\VLSR\  $\approx20$ \kms) features
were more stable.  
The systemic velocity of \vxsgr\ is estimated to be 5 \kms, 
based on a dynamical model of an expanding envelope fitted to interferometric 
maps of OH maser emission at 1612 MHz \citep{1986MNRAS.220..513C}.
Figure~\ref{fig:maser} shows the spatial distribution of \hho\ maser emission 
toward \vxsgr\ relative to the reference maser spot at \VLSR\ = 1.2 \kms\ from
observations at one of the epochs.
We found 16 maser spots detected at all epochs that could
be used for precision astrometry. In Figure~\ref{fig:maser}, 
these maser spots cluster in
eight locations identified with letters A through H in velocity order.

\begin{figure}[H]
    \centering
    \includegraphics[angle=0,scale=0.7]{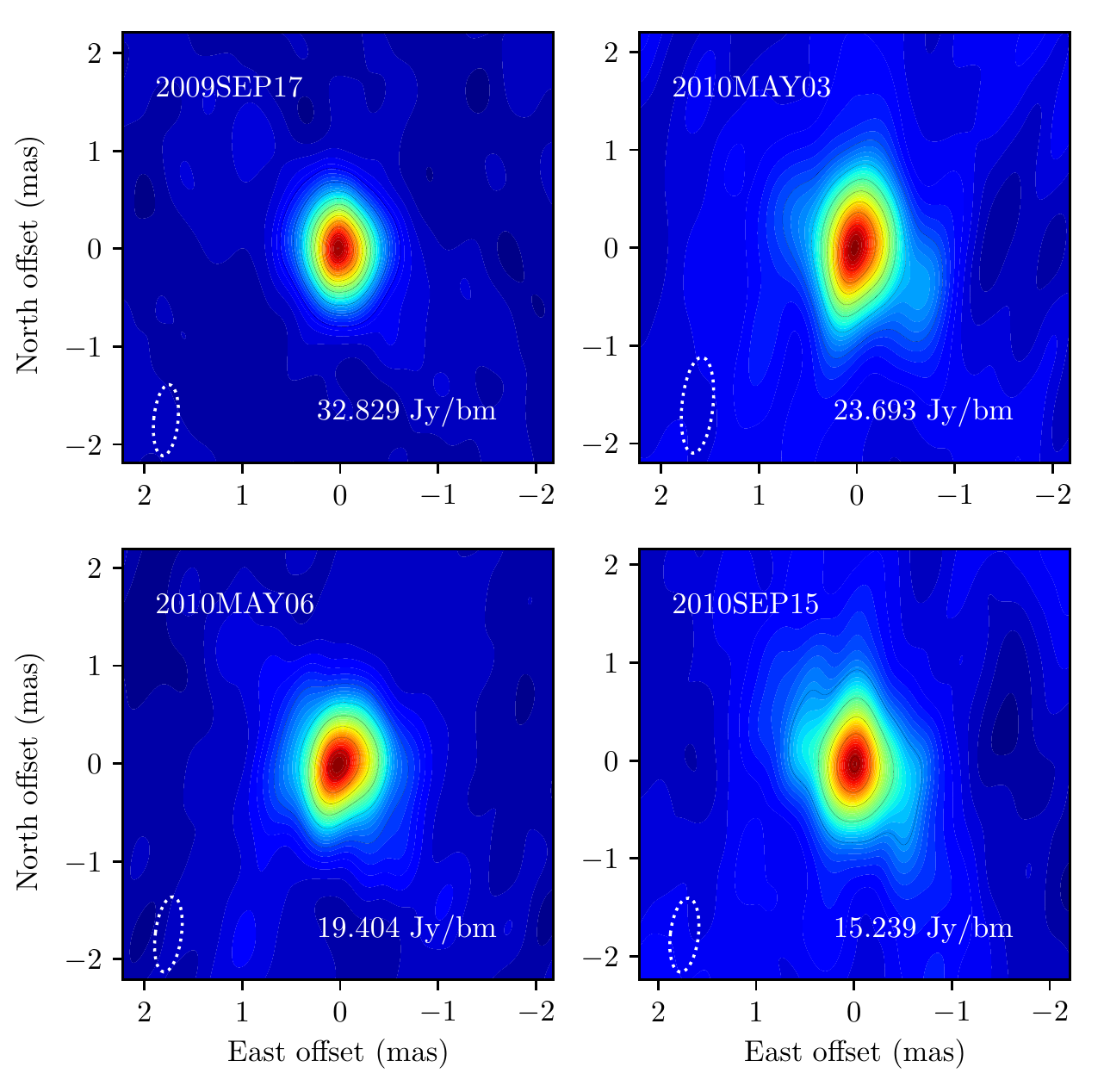}
    \caption{
    Images of the phase reference \hho\ maser spot at \VLSR\ = 1.2 \kms\ 
    toward \vxsgr\ at four epochs.
    The restoring beam is indicated in the lower left corner of each
    panel. Contour levels are spaced linearly at 3.0 \jybeam.
    }
    \label{fig:ref_ma}
\end{figure}

Figure~\ref{fig:ref_ma} shows the maser reference channel images at all
four epochs.  One can see that the emission appears dominated by a
single compact component.  Fifteen other maser spots were compact, and we 
fitted elliptical Gaussian brightness distributions to these masers 
and the extragalactic radio source \Jtwoze\ for all four epochs.  
(The extragalactic source \Jzeroei\ was not detected in our observations.)
The change in position of each maser spot relative to \Jtwoze\ 
was modeled by the parallax sinusoid in both coordinates and a 
linear proper motion in each coordinate.


We compared the astrometry results using the
different methods to calibrate residual tropospheric delays as
discussed in \S\ref{sec:trops}.
Table~\ref{tab:para_pm4} shows the parallax and proper motion fits for
the reference maser spot with different troposphere calibration methods.
All parallaxes are consistent within their joint uncertainties.
However, when the secondary calibration to remove phase bias is applied, the
parallax uncertainties decrease by about a factor of two.  While
parallax uncertainties from four epochs, a single background source, and
dominated by the East-West offsets are themselves quite uncertain 
\citep{2017AJ....154...63R}, this comparison is based on the same data 
with only a single calibration difference and should be significant.
In addition, as shown in the Figure~\ref{fig:img_qso}, the quality of the 
phase-referenced quasar images improved significantly with phase bias
corrections, providing strong evidence for the validity of this secondary calibration.

\begin{table}[H]
    \footnotesize
    \caption[]{Parallax and proper motion fits for the reference maser spot}
    \begin{center}
        \begin{tabular}{rrrrr}
            \hline \hline 
            Methods             & Parallax   &  Distance & \mux~~~~~~    & \muy~~~~~~    \\
                                & (mas)      &  (kpc)    & (\masy) & (\masy) \\
            \hline
            geodetic blocks      &  0.70 $\pm$  0.09 & 1.44 $\pm$  0.13 & 0.59 $\pm$  0.20 & -0.95 $\pm$  0.34  \\
            GPS                 &  0.56 $\pm$  0.11 & 1.78 $\pm$  0.20 & 1.42 $\pm$  0.27 &  1.81 $\pm$  1.56  \\
            geodetic blocks+bias &  0.66 $\pm$  0.05 & 1.51 $\pm$  0.07 & 0.62 $\pm$  0.10 & -0.97 $\pm$  0.25  \\
            GPS+bias            &  0.69 $\pm$  0.05 & 1.45 $\pm$  0.07 & 0.70 $\pm$  0.15 & -0.69 $\pm$  0.51  \\
            \hline
        \end{tabular}
    \end{center}
    \label{tab:para_pm4}
\end{table}

\begin{figure}[H]
  \centering
    \includegraphics[angle=0,scale=0.8]{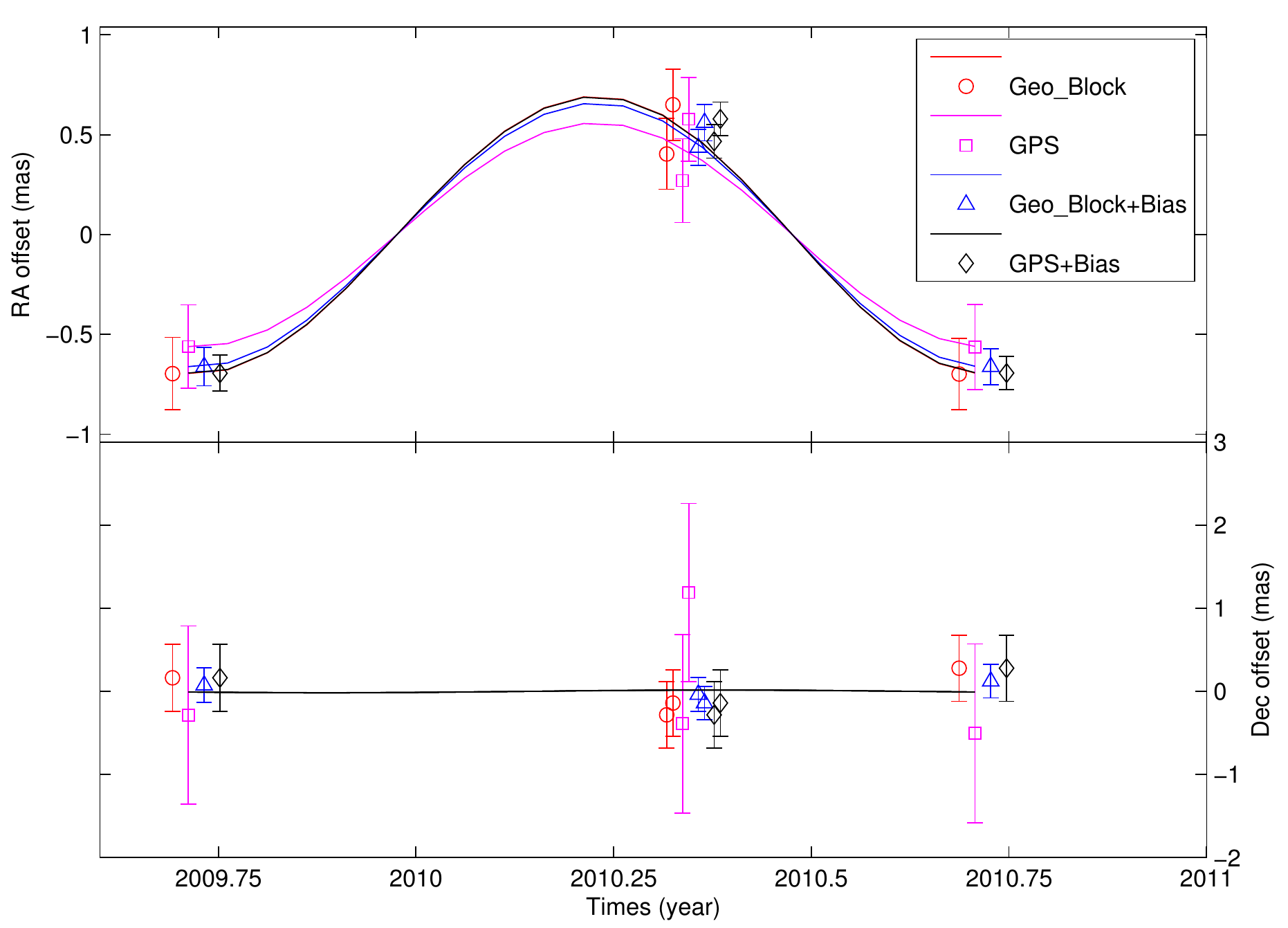}
    \caption{
Parallax and proper motion data and best-fitting models for the
reference maser spot at \VLSR\ of 1.2~\kms\ with different troposphere
calibration methods. Data for the maser spots with different methods are offset
 horizontally for clarity.}
\label{tab:para_fit4}
\end{figure}

Since one expects the same parallax for all maser spots, we did a
combined solution (fitting with a single parallax parameter for more
maser spots, but allowing for different proper motions for each maser
spot) using 16 bright maser spots.
Table~\ref{tab:com_para_fit} shows the independent and combined parallax 
fits for those maser spots.
The combined parallax estimate 
for the geodetic block plus bias corrected data is
$0.64 \pm 0.04$~mas, corresponding to a distance of
$1.56^{+0.11}_{-0.10}$~kpc.  The quoted uncertainty is the formal error
multiplied by $\sqrt{16}$ to allow for the liklihood of correlated position 
variations for the 16 maser spots.  This could result from
small variations in the background source or from mismodeled atmospheric
delays, both of which would affect the maser spots 
identically~\citep{2009ApJ...693..397R}.

\begin{table}[H]
    \footnotesize
    \caption[]{Parallax and proper motion fits for \vxsgr\ }
    \begin{center}
        \begin{tabular}{ rrrrrrrr}
            \hline \hline
         Region  &   Ch.     &  \VLSR\     &  Parallax     & \mux    & \muy     &  dx   &   dy \\ 
                 &           &  (\kms)     &   (mas)       & (\masy) & (\masy)  & (mas) & (mas)\\
            \hline
                  A  &  87 & 22.6 & 0.623 $\pm$ 0.056  &   0.58 $\pm$ 0.14  &  -0.32 $\pm$ 0.32 &  -41.17 $\pm$ 0.05 &    3.84 $\pm$ 0.11  \\ 
          A  &  89 & 21.8 & 0.668 $\pm$ 0.025  &   0.27 $\pm$ 0.07  &  -1.02 $\pm$ 0.23 &  -41.11 $\pm$ 0.02 &    3.18 $\pm$ 0.08  \\ 
          B  &  118 &  9.6 & 0.615 $\pm$ 0.065  &   0.68 $\pm$ 0.16  &  -0.95 $\pm$ 0.25 &  -20.01 $\pm$ 0.06 &    9.79 $\pm$ 0.09  \\ 
          B  &  119 &  9.2 & 0.657 $\pm$ 0.039  &   0.59 $\pm$ 0.10  &  -1.07 $\pm$ 0.54 &  -20.61 $\pm$ 0.03 &    9.18 $\pm$ 0.20  \\ 
          C  &  127 &  5.8 & 0.694 $\pm$ 0.091  &  -1.65 $\pm$ 0.22  &  -1.89 $\pm$ 0.21 & -168.01 $\pm$ 0.08 &  -23.29 $\pm$ 0.07  \\ 
          D  &  129 &  4.9 & 0.624 $\pm$ 0.046  &  -1.41 $\pm$ 0.12  &  -2.98 $\pm$ 0.32 & -135.75 $\pm$ 0.04 &  -72.10 $\pm$ 0.12  \\ 
          D  &  130 &  4.5 & 0.653 $\pm$ 0.041  &  -1.52 $\pm$ 0.10  &  -3.08 $\pm$ 0.38 & -135.87 $\pm$ 0.04 &  -71.34 $\pm$ 0.14  \\ 
          D  &  131 &  4.1 & 0.562 $\pm$ 0.017  &  -1.28 $\pm$ 0.05  &  -2.81 $\pm$ 0.34 & -136.83 $\pm$ 0.02 &  -71.08 $\pm$ 0.13  \\ 
          E  &  137 &  1.6 & 0.579 $\pm$ 0.037  &   0.56 $\pm$ 0.09  &  -1.11 $\pm$ 0.34 &  -10.29 $\pm$ 0.03 &   -1.96 $\pm$ 0.12  \\ 
          F  &  138 &  1.2 & 0.650 $\pm$ 0.031  &   0.65 $\pm$ 0.08  &  -0.97 $\pm$ 0.28 &   -0.66 $\pm$ 0.03 &    0.18 $\pm$ 0.10  \\ 
          F  &  139 &  0.7 & 0.644 $\pm$ 0.040  &   0.69 $\pm$ 0.10  &  -0.90 $\pm$ 0.25 &   -0.55 $\pm$ 0.04 &    0.31 $\pm$ 0.09  \\
          G  &  140 &  0.3 & 0.655 $\pm$ 0.036  &  -0.01 $\pm$ 0.09  &  -0.82 $\pm$ 0.38 &  -32.11 $\pm$ 0.03 &   14.67 $\pm$ 0.14  \\
          G  &  142 & -0.4 & 0.633 $\pm$ 0.037  &   0.06 $\pm$ 0.09  &  -0.78 $\pm$ 0.40 &  -31.44 $\pm$ 0.03 &   16.35 $\pm$ 0.15  \\
          H  &  158 & -7.2 & 0.650 $\pm$ 0.027  &  -1.55 $\pm$ 0.07  &  -3.51 $\pm$ 0.35 & -118.98 $\pm$ 0.03 & -110.44 $\pm$ 0.13  \\
          H  &  159 & -7.6 & 0.656 $\pm$ 0.014  &  -1.48 $\pm$ 0.04  &  -3.50 $\pm$ 0.37 & -119.02 $\pm$ 0.01 & -110.69 $\pm$ 0.14  \\
          H  &  160 & -8.0 & 0.648 $\pm$ 0.015  &  -1.50 $\pm$ 0.04  &  -3.56 $\pm$ 0.36 & -119.05 $\pm$ 0.02 & -110.87 $\pm$ 0.13  \\  
          Combined  &   &   & 0.639 $\pm$ 0.042  &    &  &   &  \\            \hline
        \end{tabular}
    \end{center}
  \tablecomments{
  Absolute proper motions are defined as
$\mux = \mu_{\alpha \cos{\delta}}$ and $\muy = \mu_{\delta}$.
  }    
    \label{tab:com_para_fit}
\end{table}

We measured the absolute proper motion of the reference maser spot (~\VLSR~ = 1.2 
\kms~) to be \mux~ = $0.65 \pm 0.08$ \masy~ and \muy~ = $-0.97 \pm 0.28$ \masy. 
In order to model the internal motions of the masers to obtain the absolute proper
 motion of the central star of \vxsgr, 
we used the 16 maser spots that appeared at four epochs within one year, and 
estimated their motions with respect to the reference maser spot. We fitted the data 
to a model of an expanding outflow.  The
estimated parameters are from a Bayesian fitting procedure described by
\citet{2010ApJ...720.1055S}.
 Converting the  systematic velocity of expanding outflow center (central star) of  
  $( V_{0x}$ = -2.19 $\pm$ 5.69 \kms,
$V_{0y}$ = -14.50 $\pm$ 5.44 \kms, $V_{0r}$ = 6.47 $\pm$ 3.37 \kms) 
to angular motions yields 
\mux\ = -0.29 $\pm$ 0.76  \masy\ and \muy\ =
-1.95 $\pm$ 0.73 \masy\ at the  distance of 1.56 kpc to \vxsgr.
Adding this motion to the absolute motion of the reference maser spot,
we estimate an absolute proper motion of \vxsgr\ to
be \mux\ = 0.36 $\pm$ 0.76 \masy\ and \muy\ = --2.92 $\pm$ 0.78 \masy.
We note that while the \HIP\ parallax (3.82 $\pm$
2.73 mas) and eastward proper motion (\mux\ = $3.20 \pm 3.08$ \masy) have
large uncertainties and allow no meaningful comparison with our results, 
their northward motion (\muy\ = $-6.71 \pm 1.72$ \masy, \citet{2007A&A...474..653V})
differs from our much more accurate result by more than $2\sigma$.
This discrepancy might be associated with inhomogeneities and
dust scattering the optical light in the circumstellar
envelope~\citep{2012ApJ...744...23Z}.

\begin{figure}[H]
  \centering
    \includegraphics[angle=0,scale=0.45]{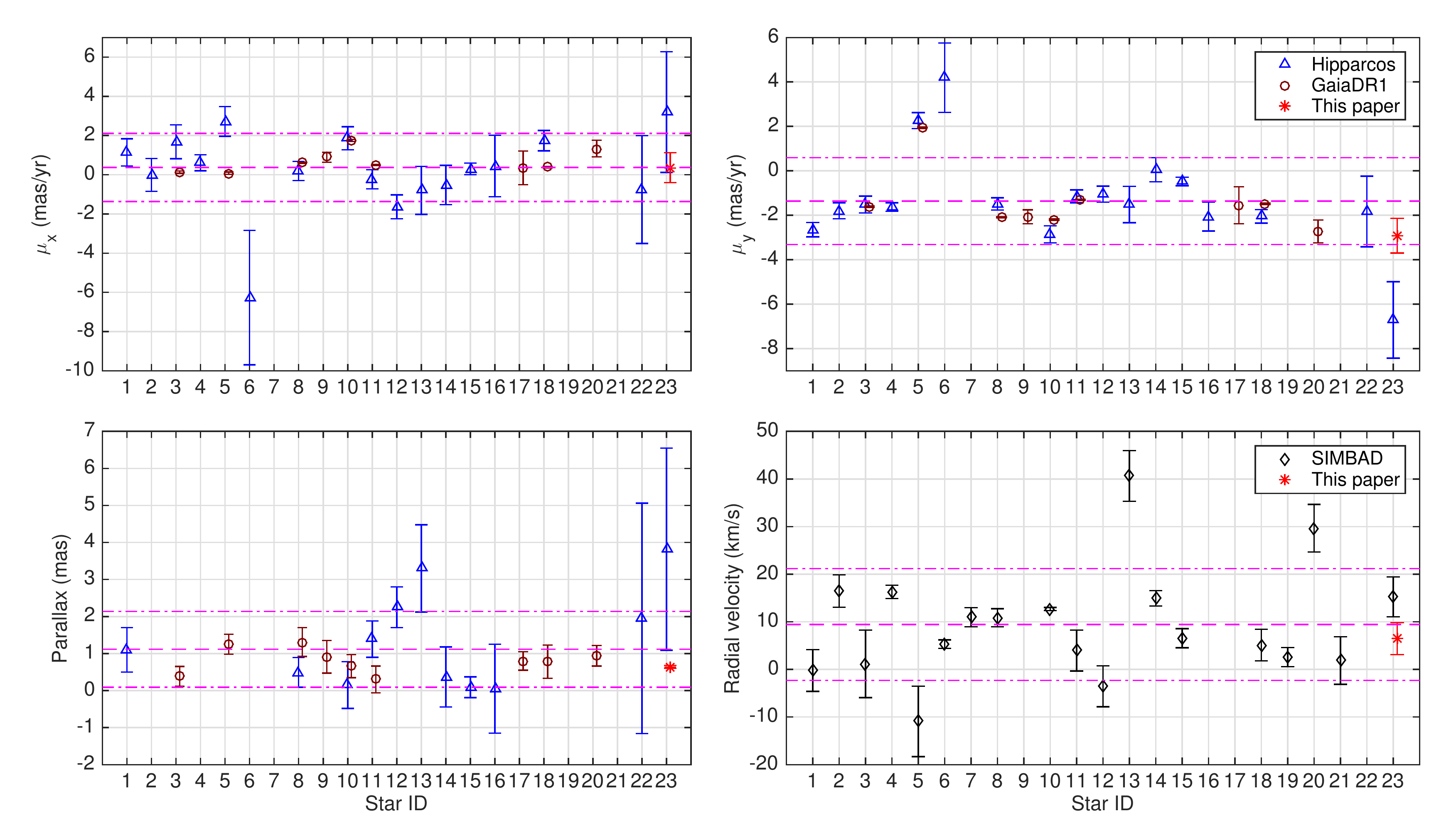}
    \caption{
Proper motion, parallax  and radial velocity results in Sgr OB1 association. 
 The star ID from 1 to 23 stands for HD 163428, HD 163800, HD 163892, HD 164402, HD 164438, HD 164492, HD 164514, HD 164794,
  HD 164816, HD 165052, HD 165516, HD 165784, HD 166167, HD 166546, HD 166937, HD 163777, HD 164018, HD 164637, HD 164883, 
  HD 164906, HD 165285, HD 166524 and \vxsgr, respectively. 
  The proper motion and parallax results are from  
  \HIP~\citep{2007A&A...474..653V}, Gaia DR1 \citep{2016A&A...595A...2G} and this paper.   The radial velocity results from the SIMBAD astronomical database \citep{2000A&AS..143....9W} are converted from heliocentric frame to the local standard of rest. 
  The horizontal dashed magenta lines indicate the means and the dash-dotted lines indicate the mean $\pm$ 1 standard deviation with the unweighted estimation for the results in Sgr OB1 association.}
\label{fig:sgrob1}
\end{figure}

Figure~\ref{fig:sgrob1} compares our astrometric
 results of \vxsgr\ with 22 of the brightest stars in the Sgr OB1 
 association \citep{1978ApJS...38..309H}. Our measured 
 astrometric results are in excellent agreement these cluster members 
 and confirm that \vxsgr\ belongs to the Sgr OB1 association.

\citet{2010A&A...511A..51C} estimated the bolometric luminosity of \vxsgr\ is
 \Lbol\ = $(0.73 \pm 0.38) \times 10^5  $ \dLsun, adopting an effective
 temperature of 3200 to 3400 K. 
  This luminosity is in good agreement with that 
 \Lbol\ = $(0.67 \pm 0.18) \times 10^5  $ \dLsun~ estimated by \citet{1982MNRAS.198..385L}, 
 from a measured magnitude at 10400 \r A and applying a bolometric correction.  
 However, \citet{2011A&A...526A.156M} determined a maximum luminosity 
 (without a quoted uncertainty) of 
 \Lbol\ = $1.38 \times 10^5  $ \dLsun, by integrating UBVIJHKL + IRAS photometry.
 \cite{2010A&A...523A..18D} derived \Lbol\ = $0.42 \times 10^5  $ \dLsun, 
 assuming the K-band magnitude is a function of pulsation period and effective temperature. 
Combined these estimated luminosity, we recalculate 
 \Lbol\ = $(0.80 \pm 0.23) \times 10^5  $ \dLsun~ using an un-weighted average and standard error of the mean (SEM) estimate, corresponding to $(1.95 \pm 0.62) \times 10^5$ \Lsun~ by adopting our 
 parallax distance of $1.56^{+0.11}_{-0.10}$ kpc.
 This luminosity coupled with an effective temperature of 3300 K suggests
 a radius for \vxsgr\ of   1,120 to 1,550 \Rsun.

\section{FUTURE OUTLOOK}
\label{sec:disc}


We have shown that for our data set there are systematic
biases of $\approx1$ cm of tropospheric zenith path-delay between GPS and VLBI 
geodetic block calibrations.   
These biases can be estimated by an image-optimization method and we 
find the primary calibration via geodetic block is slightly better than GPS
method.  Even small residual biases of $\approx1$ cm of zenith path delay can
degrade image quality and astrometric accuracy when low source elevation
data are used.  Therefore, correction of tropospheric phase-delay biases 
can be important for a very accurate parallax measurements.  The combination of
image-optimization and additional geodetic VLBI blocks or GPS 
measurements of ZWD is an effective way to accomplish this.
This method might be of great benefit when applied to archival VLBI data
taken before geodetic block calibrations were standard.

\acknowledgments

This work was partly supported by the 100 Talents Project of the Chinese Academy of Sciences, the National Science Foundation of China under grant 11673051 and 11473057, and the Key Laboratory for Radio Astronomy, Chinese Academy of Sciences.

\vspace{5mm}
\facilities{VLBA}

\software{DiFX\citep{2007PASP..119..318D}, 
          AIPS \citep{1996ASPC..101...37V}
          }

\end{CJK*}
\end{document}